\documentclass[12pt]{article}
\usepackage{graphicx,bm}

\def\Re{{\rm Re}}

\def\max{{\rm max}}

\def\norm#1{\left|\mkern-2mu\left|#1\right|\mkern-2mu\right|}
\def\norm#1{\left|\mkern-2mu\left|#1\right|\mkern-2mu\right|}

\begin{document}

\begin{center}{{\large \bf Energy dissipation and resolution of steep 
gradients in one-dimensional Burgers flows}\\~\\
Chuong V. Tran and David G. Dritschel\\
School of Mathematics and Statistics, University of St Andrews\\
St Andrews KY16 9SS, United Kingdom}
\end{center}
\date{\today}

\baselineskip=25pt


Travelling-wave solutions of the inviscid Burgers equation having smooth
initial wave profiles of suitable shapes are known to develop shocks 
(infinite gradients) in finite times. Such singular solutions are 
 characterized by energy spectra that scale with the wave number 
$k$ as $k^{-2}$. In the presence of viscosity $\nu>0$, no shocks can 
develop, and smooth solutions remain so for all times $t>0$, eventually 
decaying to zero as $t\to\infty$. At peak energy dissipation, say $t=t_*$, 
the spectrum of such a smooth solution extends to a finite 
dissipation wave number $k_\nu$ and falls off more rapidly, presumably 
exponentially, for $k>k_\nu$. The number $N$ of Fourier modes within 
the so-called inertial range is proportional to $k_\nu$. This represents 
the number of modes necessary to resolve the dissipation scale and can 
be thought of as the system's number of degrees of freedom. The peak 
energy dissipation rate $\epsilon$ remains positive and becomes 
independent of $\nu$ in the inviscid limit. 

In this study, we carry out an analysis which verifies the dynamical 
features described above and derive upper bounds for $\epsilon$ and 
$N$. It is found that $\epsilon$ satisfies 
$\epsilon \le \nu^{2\alpha-1}\norm{u_*}_\infty^{2(1-\alpha)}
\norm{(-\Delta)^{\alpha/2}u_*}^2$, where $\alpha<1$ and $u_*=u(x,t_*)$ 
is the velocity field at $t=t_*$. Given $\epsilon>0$ in the limit 
$\nu\to0$, this implies that the energy spectrum 
remains no steeper than $k^{-2}$ in that limit. For the critical 
$k^{-2}$ scaling, the bound for $\epsilon$ reduces to 
$\epsilon\le\sqrt{3}k_0\norm{u_0}_\infty\norm{u_0}^2$, where $k_0$ 
marks the lower end of the inertial range and $u_0=u(x,0)$. This 
implies $N\le\sqrt{3}L\norm{u_0}_\infty/\nu$, where $L$ is the domain 
size, which is shown to coincide with a rigorous estimate for the 
number of degrees of freedom defined in terms of local Lyapunov 
exponents. We demonstrate both analytically and numerically an 
instance where the $k^{-2}$ scaling is uniquely realizable. The 
numerics also return $\epsilon$ and $t_*$, consistent with analytic
values derived from the corresponding limiting weak solution.

\newpage

\section{Introduction}

In 1948 Burgers$^1$ introduced the equation 
\begin{eqnarray}
\label{burgers}
u_t + uu_x &=& \nu u_{xx}
\end{eqnarray}
as a model for fluid turbulence. Here, $u(x,t)$ is a one-dimensional 
velocity field and $\nu>0$ plays the role of viscosity in a usual fluid. 
On the one hand, this model captures the two most fundamental features 
of fluid dynamics by its quadratic advection and viscosity terms. On the 
other hand, Eq.\ (\ref{burgers}) lacks a pressure term, thus governing 
a hypothetical compressible fluid without pressure. The absence of a 
pressure-like term makes Eq.\ (\ref{burgers}) integrable by the Cole--Hopf 
method.$^{2,3}$ This renders Eq.\ (\ref{burgers}) and its generalization 
to higher dimensions poor models for fluid turbulence. Despite this 
apparent shortcoming, the Burgers equation has been widely studied 
for a variety of applications.$^{4-13}$

The development of shock waves or discontinuities (infinite gradients)
from suitable smooth initial velocity profiles is an intrinsic property 
of the inviscid Burgers equation. Given a differentiable initial profile 
$u(x,0)=u_0(x)$, Eq.\ (\ref{burgers}) with $\nu=0$ is implicitly solved 
by the travelling-wave solution
\begin{eqnarray}
\label{travelsolution}
u(x,t) &=& u_0(\xi) = u_0(x-ut).
\end{eqnarray}
By taking the spatial derivative of Eq.\ (\ref{travelsolution}) and 
solving the resulting equation for $u_x$ one obtains 
\begin{eqnarray}
\label{travelgradient}
u_x &=& \frac{u'_0}{1+tu'_0},
\end{eqnarray}
where $u'_0(\xi)$ denotes the derivative of $u_0(\xi)$. It follows that 
$u_x$ diverges ($u_x\to-\infty$) provided that $u'_0(\xi)<0$ for some 
$\xi$. The earliest time $t=T$ for this to occur is $T=-1/u'_0(x_0)$, 
where $u'_0(x_0)$ is the steepest slope of $u_0(x)$ occurring at $x=x_0$. 
This steepest slope travels at the speed $u_0(x_0)$ and gets ever 
steeper as $t\to T$, becoming infinitely steep when $t=T$ at 
$x=x_0+u_0(x_0)T=x_0-u_0(x_0)/u'_0(x_0)$. In summary, the space-time 
coordinate of the shock is
\begin{eqnarray}
\label{shockposition}
(x,t)=\left(x_0-\frac{u_0(x_0)}{u'_0(x_0)},\frac{-1}{u'_0(x_0)}\right).
\end{eqnarray}
Such a singular solution is characterized by an energy spectrum $E(k)$ 
that scales with the wave number $k$ as $E(k)\propto k^{-2}$, which is 
the spectrum of a step function.

Under viscous effects, the would-be shock is suppressed, and the solution
remains smooth and decays to zero in the limit $t\to\infty$. This statement
is true however small the viscosity. This means that the maximally achievable 
(peak) energy dissipation rate, hereafter denoted by $\epsilon_m$, remains 
positive in the inviscid limit $\nu\to0$. For fixed $\nu>0$, the velocity 
gradient $|u_x|$ can achieve a finite maximum only. Presumably, the 
corresponding energy spectrum would retain the $k^{-2}$ scaling up to 
a finite dissipation wave number $k_\nu$, around which the dissipation of
energy mainly takes place and beyond which a more rapid decay, probably 
exponential decay, occurs. Given this scaling, $\epsilon_m$ scales as
$\nu k_\nu$. It follows that the number $N$ of Fourier modes within the 
wave number range $k\le k_\nu$, the so-called inertial range, is 
\begin{eqnarray}
N \propto k_\nu \propto \frac{\epsilon_m}{\nu}, 
\end{eqnarray}
for dimensionally appropriate proportionality constants.
This is the number of modes necessary to resolve the dissipation scale 
and can be considered the system's number of degrees of freedom. 

In this study, we carry out an analysis that quantitatively confirms the 
dynamical features described above. It is found that $\epsilon_m$ satisfies
\begin{eqnarray}
\epsilon_m \le \nu^{2\alpha-1}\norm{u_*}_\infty^{2(1-\alpha)}
\norm{(-\Delta)^{\alpha/2}u_*}^2,
\end{eqnarray}
where $\alpha<1$, $\Delta$ is the Laplace operator, $u_*=u(x,t_*)$ is 
the velocity field at the time of peak energy dissipation $t=t_*$, and 
$\norm{\cdot}_\infty$ and $\norm{\cdot}$ denote L$^\infty$ and L$^2$ 
norms, respectively. Given that $\epsilon_m>0$ in the limit $\nu\to0$, 
this result implies that the energy spectrum $E(k,t_*)$ becomes no steeper 
than $k^{-2}$ in that limit. For this critical scaling, $\epsilon_m$ is 
found to satisfy $\epsilon_m\le\sqrt{3}k_0\norm{u_*}_\infty\norm{u_*}^2\le
\sqrt{3}k_0\norm{u_0}_\infty\norm{u_0}^2$, where $u_0=u(x,0)$ and $k_0$ is 
the wave number that marks the lower end of the energy inertial range.
This result further implies $k_\nu\le\sqrt{3}\norm{u_0}_\infty/\nu$. It
follows that $N\le\sqrt{3}L\norm{u_0}_\infty/\nu$, where $L$ is the domain
size, which is shown to coincide with a rigorous estimate for the number 
of degrees of freedom defined in terms of local Lyapunov exponents. Note 
that one can identify the upper bound for $N$ with the Reynolds number 
$\Re$ as in the case of a real fluid. Thus, the system's number of 
degrees of freedom scales linearly with $\Re$. We demonstrate both 
mathematically and numerically an instance where $E(k,t_*)\propto k^{-2}$ 
is uniquely realizable. The numerics also return the values of $\epsilon_m$ 
and $t_*$ which are consistent with those derived from the corresponding 
limiting weak solution. 

\section{Energy dissipation and dissipaton wave number}

For simplicity, we consider periodic solutions of Eq.\ (\ref{burgers}) 
having period $2\pi L$ and vanishing spatial average. The usual L$^p$ 
norm of $u$ (and of its derivatives), for all $p>0$ including $p=\infty$, 
is defined by $\norm{u}_p=\langle |u|^p\rangle^{1/p}$, where 
$\langle\cdot\rangle$ denotes a domain average. The advection term of 
the Burgers equation conserves $\norm{u}_p$. Under viscous effects, 
$\norm{u}_p$ decays for $p\ge1$ and is governed by. 
\begin{eqnarray}
\label{norm-decay}
\frac{d}{dt}\norm{u}_p &=& 
-\nu(p-1)\norm{u}_p^{1-p}\langle|u|^{p-2}u_x^2\rangle.
\end{eqnarray}
Since we are dealing with L$^2$ 
and L$^\infty$ norms only, we omit the subscript $p=2$ in the former
for convenience. The decay of the energy $\norm{u}^2/2$ is governed by
\begin{eqnarray}
\frac{1}{2}\frac{d}{dt}\norm{u}^2 &=& -\nu\norm{u_x}^2.
\end{eqnarray}
This section is mainly interested in optimal estimates for the decay 
rate $\nu\norm{u_x}^2$, particularly in the limit of small $\nu$, and 
related issues concerning the energy inertial range.

The governing equation for the velocity gradient $u_x$ is
\begin{eqnarray}
\label{grad}
u_{xt} + uu_{xx}+u_x^2 &=& \nu u_{xxx}.
\end{eqnarray}
By multiplying Eq.\ (\ref{burgers}) by $u_{xx}$ (or Eq.\ (\ref{grad}) by 
$u_{x}$) and integrating the resulting equation over the domain we obtain 
the evolution equation for the mean-square velocity gradient $\norm{u_x}^2$,
\begin{eqnarray}
\label{gradient}
\frac{1}{2}\frac{d}{dt}\norm{u_x}^2 
&=& 
\langle u_{xx}uu_x\rangle - \nu\norm{u_{xx}}^2 \nonumber\\
&\le& 
\norm{u}_\infty\norm{u_x}\norm{u_{xx}} - \nu\norm{u_{xx}}^2 \nonumber\\
&=& 
\frac{\norm{u_{xx}}^2}{\norm{u_x}^2}\left(\norm{u}_\infty
\frac{\norm{u_x}^3}{\norm{u_{xx}}} - \nu\norm{u_x}^2\right)
\end{eqnarray}
where the inequality is straightforward. The final line of Eq.\ 
(\ref{gradient}) can be used to derive an upper bound for the energy 
dissipation rate $\nu\norm{u_x}^2$. For this purpose, consider the 
inequality (see Eq.\ (7) of Ref.\ 14) 
\begin{eqnarray}
\label{holder}
\frac{\norm{u_x}^3}{\norm{u_{xx}}} &\le& 
\frac{\norm{(-\Delta)^{\alpha/2}u}^{1/(1-\alpha)}}
{\norm{u_x}^{(2\alpha-1)/(1-\alpha)}},
\end{eqnarray}
where $\alpha<1$ is a parameter, which can be varied for an optimal 
bound, and $\Delta$ is the Laplace operator. The fractional derivative 
$(-\Delta)^{\alpha/2}$ is a positive operator and is defined by
$\widehat{(-\Delta)^{\alpha/2}u}=k^\alpha\widehat{u}$, where 
$\widehat{(-\Delta)^{\alpha/2}u}$ and $\widehat{u}$ are the Fourier 
transforms of $(-\Delta)^{\alpha/2}u$ and $u$, respectively. Upon 
substituting Eq.\ (\ref{holder}) into Eq.\ (\ref{gradient}) and 
noting that $d\norm{u_x}^2/dt=0$ at the time of peak energy 
dissipation $t=t_*$, we can deduce that 
\begin{eqnarray}
\label{epsilonbound}
\epsilon_m &\le& \nu^{2\alpha-1}\norm{u_*}_\infty^{2(1-\alpha)}
\norm{(-\Delta)^{\alpha/2}u_*}^2,
\end{eqnarray}
where $\norm{u_*}_\infty$ is bounded by its initial value, but
$\norm{(-\Delta)^{\alpha/2}u_*}$ can be large, depending on both 
$E(k,t_*)$ and $\alpha$. In section IV, we demonstrate both 
analytically and numerically that in the limit $\nu\to0$, $t^*$ 
is independent of $\nu$ and, in general, not related to the 
singularity time $T$ of the corresponding inviscid solution. 

Equation (\ref{epsilonbound}) confirms the fact that $\epsilon_m<\infty$ 
(and hence $\norm{u_x}<\infty$) for $\nu>0$ as one can set $\alpha=0$ 
and obtain $\epsilon_m\le\norm{u_*}_\infty^2\norm{u_*}^2/\nu\le\norm{u_0}_
\infty^2\norm{u_0}^2/\nu$. This bound can be highly excessive, and a more 
optimal estimate is possible by varying the ``optimization'' parameter 
$\alpha$ within the permissible range $\alpha<1$. Observe that the 
spectrum of $\norm{(-\Delta)^{\alpha/2}u}^2/2$ is $k^{2\alpha}E(k)$. 
So, if the energy spectrum $E(k,t_*)$ is strictly steeper than $k^{-2}$, 
then $\norm{(-\Delta)^{\alpha/2}u_*}$ is bounded for some $\alpha>1/2$. 
If this were the case for all $\nu$, including the limit $\nu\to0$, 
then the upper bound for $\epsilon_m$ in Eq.\ (\ref{epsilonbound}) would 
vanish, thereby contradicting the fact that $\epsilon_m>0$ in that limit. 
This rules out energy spectra steeper than $k^{-2}$. In section IV,
we mathematically demonstrate an instance where energy spectra shallower 
than $k^{-2}$ are also ruled out. Thus the scaling $k^{-2}$ is uniquely
realizable. This suggests that in general, the most plausible scenario 
is that in the inviscid limit, $E(k,t_*)$ approaches the $k^{-2}$ 
critical scaling. 

Now, suppose that $E(k)=Ck^{-2}/2$, for $k\in[k_0,k_\nu]$, where $C>0$ is 
a constant. Note that $k_0$ is not necessarily the lowest wave number $1/L$.
We then have $\norm{u}^2=C\int_{k_0}^{k_\nu}k^{-2}\,dk$, so 
$C=k_0\norm{u}^2$. Thus, $E(k)=k_0\norm{u}^2k^{-2}/2$. For this 
case, a direct estimate of the ratio $\norm{u_x}^3/\norm{u_{xx}}$ is
\begin{eqnarray}
\frac{\norm{u_x}^3}{\norm{u_{xx}}} &=& \sqrt{3}k_0\norm{u}^2.
\end{eqnarray}
By applying this equation to $u_*$ and substituting the resulting 
estimate into Eq.\ (\ref{gradient}) we deduce the upper bound
\begin{eqnarray}
\label{epsilonbound1}
\epsilon_m &\le& \sqrt{3}k_0\norm{u_*}_\infty\norm{u_*}^2
\le \sqrt{3}k_0\norm{u_0}_\infty\norm{u_0}^2.
\end{eqnarray}
We find later by an example for the parameter values $\norm{u_0}_\infty=1$, 
$\norm{u_0}^2=1/2$ and $k_0=1$ that $\epsilon_m=0.1061$,
which gives us a sense of the sharpness of the derived upper bound 
$\sqrt{3}k_0\norm{u_0}_\infty\norm{u_0}^2=\sqrt{3}/2$. The 
dissipation wave number $k_\nu$, which marks the end of the $k^{-2}$ 
inertial range, is found to satisfy
\begin{eqnarray}
\label{knubound}
k_\nu &\le& \frac{\sqrt{3}\norm{u_0}_\infty}{\nu}.
\end{eqnarray}
It follows that the number $N$ of Fourier modes within this inertial 
range is bounded by
\begin{eqnarray}
\label{Nbound}
N &\le& \frac{\sqrt{3}L\norm{u_0}_\infty}{\nu} = \Re,
\end{eqnarray}
where $\Re$ is the Reynolds number. Note that this estimate also includes
the modes corresponding to $k<k_0$. The linear dependence of $N$ on $\Re$ 
is interesting and is rigorously verified, without reference to 
$E(k,t_*)$, in what follows. 
 
\section{Lyapunov exponents and number of degrees of freedom}
 
This section derives a rigorous estimate for the number of degrees of 
freedom, which is defined as the minimum number of greatest local Lyapunov 
exponents (of a general trajectory in phase space) whose sum becomes 
negative. This number, denoted by $D$, is the dimension of the linear 
space (spanned by the corresponding Lyapunov vectors), which can adequately 
``accommodate'' the solution locally, and is essentially the so-called 
Lyapunov or Kaplan--Yorke dimension.$^{15,16}$ 
Its estimate is found to agree with that for 
$N$ obtained earlier in the preceding section. This agreement is not 
coincidental and can be considered as analytic evidence for the expected 
$k^{-2}$ energy spectrum used in the estimation of $N$. Like $N$, $D$ 
can be thought of as the number of Fourier modes necessary to resolve 
the steepest velocity gradient during the course of evolution, particularly 
around $t=t_*$. We follow the procedure formulated by Tran and 
Blackbourn$^{17}$ in the calculation of the number of degrees of freedom 
for two-dimensional Navier--Stokes turbulence. For a detailed discussion of 
the significance of $D$, see Refs.\ 17 and 18 and references therein.
 
Given the solution $u(x,t)$ starting from some smooth initial velocity 
field $u_0(x)$, consider a disturbance $v(x,t)$ satisfying the same
conditions as $u(x,t)$, i.e., periodic boundary condition and 
zero spatial average. The linear evolution of $v(x,t)$ is governed by
\begin{eqnarray}
\label{linearized}
v_t + uv_x + vu_x&=& \nu v_{xx}.
\end{eqnarray}
The governing equation for the norm $\norm{v}$ is
\begin{eqnarray}
\label{vnorm}
\norm{v}\frac{d}{dt}\norm{v} &=& -\langle v(uv_x + vu_x)\rangle - 
\nu \norm{v_x}^2 \nonumber\\
&=& \langle uvv_x\rangle - \nu \norm{v_x}^2 \nonumber\\
&\le& \norm{u}_\infty\norm{v}\norm{v_x} - \nu \norm{v_x}^2 \nonumber\\
&\le& \norm{u_0}_\infty\norm{v}\norm{v_x} - \nu \norm{v_x}^2,
\end{eqnarray}
where we have used $\langle v^2u_x\rangle=-2\langle uvv_x\rangle$ by 
integration by parts and the inequalities are straightforward. Dividing
both sides of Eq.\ (\ref{vnorm}) by $\norm{v}^2$ yields
\begin{eqnarray}
\label{lambda}
\lambda = \frac{1}{\norm{v}}\frac{d}{dt}\norm{v}
&\le& \norm{u_0}_\infty\frac{\norm{v_x}}{\norm{v}} 
- \nu \frac{\norm{v_x}^2}{\norm{v}^2},
\end{eqnarray}
where $\lambda$ is the exponential rate of growth ($\lambda>0$) or decay 
($\lambda<0$) of the disturbance norm $\norm{v}$.

The set of $n$ greatest local Lyapunov exponents 
$\{\lambda_1,\lambda_2,\cdots,\lambda_n\}$ and the corresponding orthonormal
set of $n$ most unstable disturbances $\{v^1,v^2,\cdots,v^n\}$ can be 
derived by successively maximizing $\lambda$ with respect to all admissible 
disturbances $v$ subject to the following orthogonality constraint. At each
step $i$ in the process, the maximizer $v$ is required to satisfy both 
$\norm{v}=1$ and $\langle vv^j\rangle=0$, for $j=1,2,\cdots,i-1$, where 
$v^j$ is the solution obtained at the $j$-th step. Since each normalized 
solution $(\lambda_i,v^i)$ satisfies Eq.\ (\ref{lambda}), we have
\begin{eqnarray}
\label{lambda-sum}
\sum_{i=1}^n\lambda_i 
&\le& 
\norm{u_0}_\infty\sum_{i=1}^n\norm{v^i_x} - \nu\sum_{i=1}^n\norm{v^i_x}^2
\nonumber\\
&\le& 
\norm{u_0}_\infty\left(n\sum_{i=1}^n\norm{v^i_x}^2\right)^{1/2} 
- \nu\sum_{i=1}^n\norm{v^i_x}^2
\nonumber\\
&=& 
\left(\sum_{i=1}^n\norm{v^i_x}^2\right)^{1/2}\left(\norm{u_0}_\infty n^{1/2} 
- \nu\left(\sum_{i=1}^n\norm{v^i_x}^2\right)^{1/2}\right)
\nonumber\\
&\le& 
\left(n\sum_{i=1}^n\norm{v^i_x}^2\right)^{1/2}\left(\norm{u_0}_\infty 
- \frac{\nu n}{cL}\right),
\end{eqnarray}
where $c$ is a constant independent of the orthonormal set in question.
In Eq.\ (\ref{lambda-sum}), we have applied the Cauchy--Schwarz inequality 
$\sum_{i=1}^n\norm{v^i_x}\le(n\sum_{i=1}^n\norm{v^i_x}^2)^{1/2}$ 
and used the estimate
\begin{eqnarray}
\label{rayleigh}
\sum_{i=1}^n\norm{v^i_x}^2 &\ge& \frac{n^3}{c^2L^2},
\end{eqnarray}
which is a consequence of the Rayleigh--Ritz principle. By this
principle, the left-hand side of Eq.\ (\ref{rayleigh}) is not smaller 
than the sum of the first (i.e., smallest) $n$ eigenvalues of $-\Delta$.
These eigenvalues are $1/L^2,2^2/L^2,\cdots,n^2/L^2$ and sum up to 
$n(n+1)(2n+1)/(6L^2)$. Hence, Eq.\ (\ref{rayleigh}) follows with $c$ tending
to $\sqrt{3}$ for large $n$. Now the condition $\sum_{i=1}^n\lambda_i\le0$ 
is satisfied when $n\ge cL\norm{u_0}_\infty/\nu$. It follows that
\begin{eqnarray}
\label{degree}
D &\le& c\,\frac{L\norm{u_0}_\infty}{\nu}.
\end{eqnarray}  
This estimate agrees with the upper bound (\ref{Nbound}) for $N$, which 
was derived by assuming the energy spectrum $E(k)\propto k^{-2}$. This
agreement provides us with confidence in the plausibility of the $k^{-2}$ 
scaling.  

The term on the right-hand side of Eq.\ (\ref{degree}) is the Reynolds 
number $\Re$ defined earlier with $c=\sqrt{3}$. Thus $D$ scales linearly 
with $\Re$. For a comparison, $D$ scales as $\Re(1+\ln\Re)^{1/3}$ and 
$\Re^{9/4}$ for two-dimensional and three-dimensional turbulence, 
respectively. The former has recently been derived$^{17}$ while the 
latter is a classical result deduced from the Kolmogorov theory. These 
scalings reflect the intrinsic characteristics that the dynamics of the 
two-dimensional vorticity gradient and three-dimensional vorticity are 
effectively linear and quadratically nonlinear, respectively.$^{18,19}$ The 
present finding of exactly linear dependence of $D$ on $\Re$ is somewhat 
unexpected as the Burgers velocity gradient dynamics are quadratically 
nonlinear, just as the three-dimensional vorticity dynamics. Nonetheless, 
this is not a total surprise if the dimension of the physical space, which 
plays a significant role in the scaling of $D$ with $\Re$, is taken into 
account.$^{18}$ Note that in all three cases, $D$ scales linearly with 
the domain volume, given all else fixed. This is in accord with the notion 
of extensive chaos.$^{20-22}$ The linear scaling of $D$ with $\Re$ for 
the Burgers case is fully justified in the numerical simulations reported 
in the next section, where we observe that the ratio $D/\Re$ is best kept 
fixed (at order unity) for various resolutions. Hence, doubling the 
resolution (i.e., doubling $D$) allows the viscosity to be halved, given 
all else fixed. This allows the exponential dissipation rate $\nu k^2$ at 
the truncation wave number to grow as $\Re$. On the other hand, this same 
linear scaling of $D$ with $\Re$ in two-dimensional turbulence means that 
numerical simulations can be performed using a fixed dissipation rate 
$\nu k^2$ at the truncation wave number, for different resolutions. Thus, 
doubling the resolution (i.e., quadrupling $D$) allows the viscosity to be 
reduced by the factor $1/4$. This fact is well known to numerical analysts. 
The scaling of $D$ as $\Re^{9/4}$ in three-dimensional turbulence implies 
that the dissipation rate $\nu k^2$ at the truncation wave number should 
be proportional to $\Re^{1/2}$. This means that doubling the resolution 
(i.e., octupling $D$) can allow the viscosity to be reduced by the factor 
$2^{-4/3}$.

\section{A case study}

In this section we analytically and numerically consider an example 
that confirms the results derived in the preceding sections. In 
addition, we prove that no power-law energy spectra other than $k^{-2}$ 
are realizable, thus giving an exact result of the slope of $E(k,t_*)$
rather than a constraint for this particular case. We also determine 
by numerical simulations the viscosity-independent maximum dissipation 
rate $\epsilon_m$ and the corresponding time $t=t_*$ when this occurs. 
The numerical values of these dynamical parameters agree with those
derived from the corresponding limiting weak solution.

\subsection{Analytical consideration}

\begin{figure} 
\begin{center}
\includegraphics[width=12.5cm]{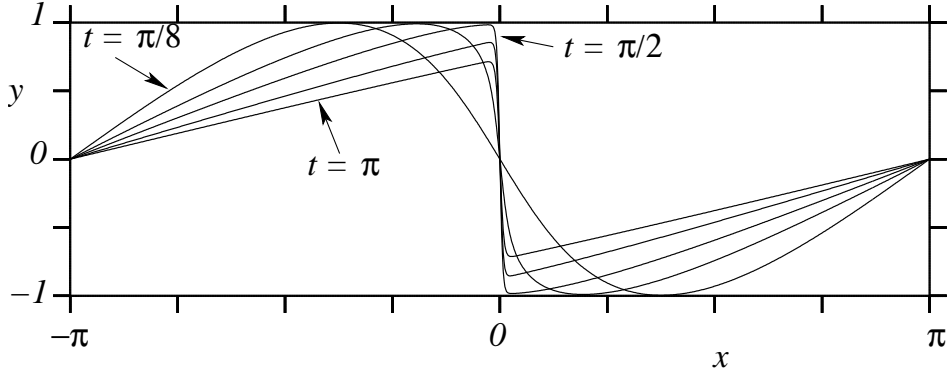}
\end{center}
\caption{A viscous solution to Burgers equation starting from
$u(x,0)=-\sin x$, for $\nu = 0.02$, and 
shown at times $t=\pi/8$, $5\pi/16 \approx 1$, $\pi/2$, $3\pi/4$ 
and $\pi$.}
\label{fig1}
\end{figure}

We consider the periodic domain $[-\pi,\pi]$, i.e., $L=1$, and 
$u_0(x)=-\sin x$. This initial profile was used in a computational 
study$^{13}$ of the Burgers equation, using 4096 grid points. In the 
next subsection, we report results from simulations
using up to $4\times10^4$ Fourier modes. It can be readily seen 
that Eq.\ (\ref{burgers}) admits odd functions as solutions. In 
other words, if $f(x,t)$ is a solution, then $f(-x,t)$ is also a 
solution provided that $f(x,t)=-f(-x,t)$. Hence, for the initial 
profile under consideration, $u(x,t)$ remains odd for all $t>0$. 
We can then express $u(x,t)$ in terms of an odd Fourier series:
\begin{eqnarray}
\label{fourier}
u(x,t) &=& \sum_k u_k(t)\sin kx,
\end{eqnarray}
where $k=1,2,3,\cdots$ are the wave numbers. The gradient $u_x$ is given by
\begin{eqnarray}
\label{gradient1}
u_x(x,t) &=& \sum_k ku_k(t)\cos kx.
\end{eqnarray}
The origin is ``stationary'' and has the steepest negative slope, 
initially equalling $-1$, which is given in terms of $u_k$ by
\begin{eqnarray}
\label{gradient2}
u_x(0,t) &=& \sum_k ku_k(t).
\end{eqnarray}
The third derivative $u_{xxx}(0,t)$ is
\begin{eqnarray}
\label{Gradient}
u_{xxx}(0,t)=-\sum_kk^3u_k(t). 
\end{eqnarray}
By substituting Eqs.\ (\ref{gradient2}) and (\ref{Gradient})
into Eq.\ (\ref{grad}) one obtains 
\begin{eqnarray}
\label{gradient3}
\frac{\partial}{\partial t}\sum_k ku_k &=& -\left(\sum_k ku_k\right)^2
-\nu\sum_k k^3u_k.
\end{eqnarray} 
In the inviscid case, $u_x(0,t)\to-\infty$ as $t\to T=1$. This can be 
seen either by solving Eq.\ (\ref{gradient3}) with $\nu=0$ or directly
from Eq.\ (\ref{shockposition}). Figure \ref{fig1} illustrates 
the viscous solution (for $\nu = 0.02$) at a few selected times before, 
near and after the inviscid singularity time ($t = 1$).

The evolution of the Fourier coefficients $u_k(t)$ is governed by
\begin{eqnarray}
\label{uk}
\frac{\partial}{\partial t} u_k &=& \frac{k}{4}u_{k/2}^2\mp\frac{k}{2}
\sum_{m\pm\ell=k} u_mu_\ell - \nu k^2u_k,
\end{eqnarray}
where the sum is over all pairs of wave numbers $m$ and $\ell$, including
$m=\ell=k/2$ when $k$ is even, satisfying the triad condition $m\pm\ell=k$. 
Within each individual wave number triad, the energy is conservatively
transferred from each of the two lower wave numbers to the third and higher 
wave number or vice versa. It can be seen that all wave numbers are 
initially excited in such a way that $u_k<0$. Plausibly, no particular 
modes would become completely depleted of energy during the subsequent 
evolution. This means that $u_k$ does not change sign and remains negative. 
This fact is verified below in the numerical simulations. As a consequence, 
the transfer of energy to ever-smaller scales is irreversible, and each 
Fourier mode contributes to the steepness of the slope $u_x(0,t)$ as 
there are no cancellations in the sum $\sum_k ku_k$. The nonlinearity 
can be said to operate at ``full strength,'' without ``depletion.'' This 
is consistent with the fact that $u_x(0,t)$ quickly diverges if $\nu=0$;
indeed $u_x(0,1)=-\infty$. This observation prompts us to take $u_k<0$ 
for all $k$ in what follows.

Consider the inertial range scaling $u_k=-c_\gamma k^{-\gamma}$, for 
$0<\gamma<3/2$ and $c_\gamma>0$, which corresponds to the energy 
spectrum $E(k)=c_\gamma^2k^{-2\gamma}/2$. 
By substituting this scaling for $u_k$ into the right-hand side of Eq.\
(\ref{gradient3}) we obtain
\begin{eqnarray}
\label{gradient4}
\frac{\partial}{\partial t}\sum_k ku_k 
&=& 
-\frac{c_\gamma^2k_\nu^{4-2\gamma}}{(2-\gamma)^2}
+\nu\frac{c_\gamma k_\nu^{4-\gamma} }{4-\gamma} \nonumber\\
 &=& 
k_\nu^{1+\gamma}\left(\frac{(3-2\gamma)\epsilon_m}{c_\gamma(4-\gamma)}
-\frac{c_\gamma^2k_\nu^{3-3\gamma}}{(2-\gamma)^2}\right)
\end{eqnarray} 
where $\epsilon_m=\nu c_\gamma^2k_\nu^{3-2\gamma}/(3-2\gamma)$ has been 
calculated from the above spectrum. The fact that both 
$\sum_k ku_k\to-\infty$ and $0<\epsilon_m<\infty$ as $k_\nu\to\infty$
requires $\gamma=1$, which is the only possibility allowed by Eq.\ 
(\ref{gradient4}). Indeed, if $\gamma>1$ (which has already been ruled 
out in general), then the second term in the brackets of Eq.\ 
(\ref{gradient4}) could be made arbitrarily small for sufficiently 
large $k_\nu$ and the right-hand side would become positive. This
contradicts the fact that $\sum_k ku_k\to-\infty$. On the other hand, 
if $\gamma<1$, then the second term in the brackets of Eq.\ 
(\ref{gradient4}) could be made arbitrarily large for sufficiently 
large $k_\nu$ and the right-hand side would become negative. The
gradient at the origin $\sum_k ku_k$ would diverge for $k_\nu<\infty$,
which is not possible.

\begin{figure} 
\begin{center}
\includegraphics*[5cm,13.5cm][20cm,21cm]{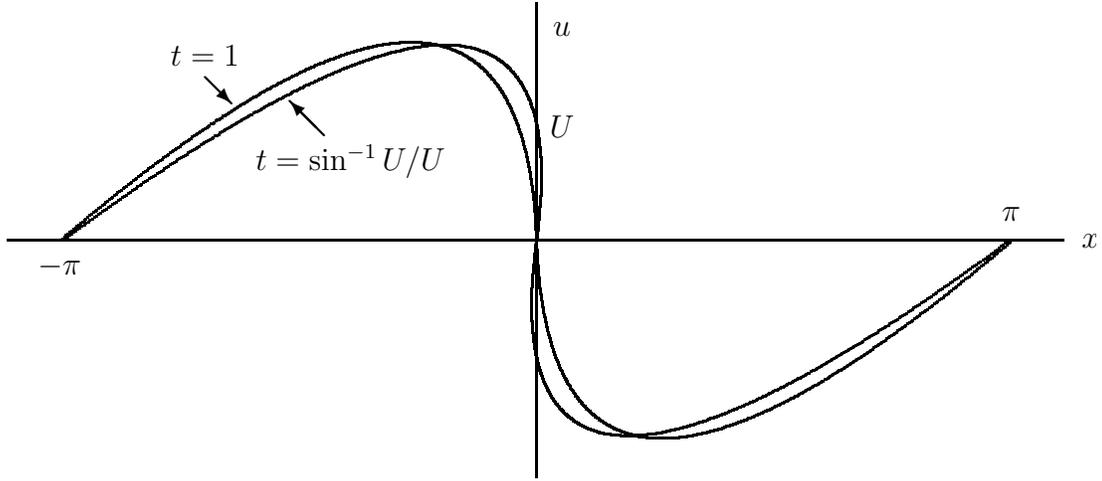}
\end{center}
\caption{A schematic description of energy loss after wave 
breaking at $t=1$ for the travelling-wave solution $u=-\sin(x-ut)$ of the 
inviscid Burgers equation. The energy dissipation rate is $U^3(t)/(3\pi)$, 
where $2U(t)$ is the shock width. This rate is zero upon wave breaking and 
grows to its maximum of $1/(3\pi)$ at $t=\pi/2$.}
\label{cartoon}
\end{figure}

We now consider the energy dissipation rate in the inviscid case due to 
the lack of smoothness of solution after wave breaking at $t=1$. This
consideration allows us to determine the energy dissipation rate, among
other things, of the viscous case in the inviscid limit. For $t>1$, the 
travelling-wave solution becomes multivalued in a neighborhood of $x=0$ 
as the respective portions $u>0$ and $u<0$ of $u$ cross over the vertical 
axis, invading the region $x>0$ and $x<0$ (see Figure \ref{cartoon}). 
Consider the weak solution consisting of two disconnected travelling-wave 
branches $u_+(x,t)$ and $u_-(x,t)$ given by
\begin{eqnarray}
u_+(x,t) &=& \cases{
\begin{array}{ccc}
-\sin(x-u_+t) & \mbox{~for~} & -\pi \le x \le 0 \\
0 & \mbox{~for~} & 0 < x \le \pi
\end{array} \cr}
\end{eqnarray}
and
\begin{eqnarray}
u_-(x,t) &=& \cases{
\begin{array}{ccc}
-\sin(x-u_-t) & \mbox{~for~} & 0 \le x \le \pi \\
0 & \mbox{~for~} & -\pi \le x < 0. 
\end{array} \cr}
\end{eqnarray}
These terminate on the vertical axis at $u_+(0,t)=U(t)$ and $u_-(0,t)=-U(t)$,
where the (half) shock width $U(t)$ is given implicitly by $U=\sin(Ut)$. 
Evidently, $\lim_{t\to1_+}U(t)=0$ and $U(\pi/2)=1$, the latter of which 
is the global maximum. The evolution of the energy corresponding to this 
solution is governed by
\begin{eqnarray}
\label{erate}
\frac{1}{2}\frac{d}{dt}\norm{u}^2 
&=&
-\frac{1}{2\pi}\left(\int_{-\pi}^0u_+^2(u_+)_x\,dx +
\int_0^\pi u_-^2(u_-)_x\,dx \right) \nonumber\\
&=&
-\frac{1}{6\pi}\left(\int_{-\pi}^0(u_+^3)_x\,dx +
\int_0^\pi(u_-^3)_x\,dx \right) \nonumber\\
&=&
-\frac{U^3}{3\pi}.
\end{eqnarray}
The energy dissipation rate $U^3/(3\pi)$ tends to zero as $t\to1_+$ and 
achieves its maximum of $1/(3\pi)$ at $t=\pi/2$ when $U(\pi/2)=1$. For
$t>\pi/2$, this rate decreases monotonically to zero as $t\to\infty$.
Since the viscous solution approaches this (unique) weak solution in 
the limit $\nu\to0$, the limiting energy dissipation rate for $t\ge1$ 
is $U^3/(3\pi)$. The maximum dissipation rate corresponds to $U=1$, i.e., 
$\epsilon_m=1/(3\pi)$, occurring at $t=t_*=\pi/2$. Note that $t_*$ differs
from $T$ and is the time for the extrema (initially at $x=\pm\pi/2$) 
to arrive at the stationary shock position $x=0$. In the next subsection,
we recover both values of $\epsilon_m$ and $t_*$ with high precision by 
numerical simulations.

An interesting feature of the present problem is that in the inviscid 
limit the energy commences its decay from $t=1$ while the maximum velocity 
does so from $t=\pi/2$, upon which the energy dissipation reaches its peak.
This lag in the dissipation of $\norm{u}_\infty$ can be readily appreciated
by the following observation. For the energy, the dissipation rate is 
dominated by $|u_x(0,t)|$, which becomes sufficiently large at $t=1$,
upon which the transition between nondissipative and dissipative phases 
takes place. For the maximum velocity, by taking the limit $p\to\infty$ 
of Eq.\ (\ref{norm-decay}) we obtain
\begin{eqnarray}
\label{norm-infty1}
\frac{d}{dt}\norm{u}_\infty &=& 
-\nu\lim_{p\to\infty}(p-1)\norm{u}_p^{1-p}\langle|u|^{p-2}u_x^2\rangle.
\end{eqnarray}
The dissipation rate on the right-hand side of Eq.\ (\ref{norm-infty1})
is dominated by $|u_x|$ in the vicinity of the maximum velocity. Evidently,
as the maximum velocity approaches the vertical axis, $|u_x|$ in its vicinity
becomes greater (see figure \ref{fig1}). The transition between 
inviscid and viscous dynamics of $\norm{u}_\infty$ at $t=\pi/2$ implies that 
$|u_x|$ in this vicinity is not sufficiently large until $t=\pi/2$. A 
similar behavior has been observed numerically in two-dimensional 
turbulence, whereby the vorticity supremum remains virtually unchanged 
until (and even after) the dissipation rate of the mean square vorticity 
has achieved its maximum value.$^{23}$ 

The weak solution provides a convenient way for calculating the 
dissipation rate $d\norm{u}_\infty/dt$ for $t\ge\pi/2$. In the limit 
$\nu\to0$, one can identify $\norm{u}_\infty$ with $U=\sin(Ut)$. By 
taking the time derivative of this expression and solving the resulting 
equation for $dU/dt=d\norm{u}_\infty/dt$ we obtain
\begin{eqnarray}
\label{norm-infty2}
\frac{d}{dt}\norm{u}_\infty &=& 
-\frac{\norm{u}_\infty(1-\norm{u}_\infty^2)^{1/2}}
{1+t(1-\norm{u}_\infty^2)^{1/2}}.
\end{eqnarray}

In the present example, $-u_x(x,0)$ peaks at an isolated point, namely at 
$x=0$. The weak solution is a step function with $U(T)=0$ and the energy
dissipation rate tends to zero as $t\to T_+$. Similarly, consider a 
smooth initial profile $u(x,0)$, for which $-u_x(x,0)$ achieves a 
positive maximum at a finite number, say $N_0$, of isolated points. 
Such a profile evolves into a piecewise smooth solution having $N_0$ 
steps, each with $U(T)=0$. For this case, the energy dissipation rate 
also tends to zero as $t\to T_+$. When the said maximum occurs over an 
extended interval, say $[x_1,x_2]$, then $U(T)=(x_1-x_2)u_x(x_1,0)>0$. 
The energy dissipation rate upon wave breaking jumps from zero to a 
positive value.

\subsection{Numerical results}

We now turn to results of a numerical analysis of the Burgers equation. 
We have simulated the initial value problem described by Eq.\ (\ref{uk}), 
where $u_1(0)=-1$ and $u_k(0)=0$ for $k>1$, for several different resolutions 
up to $k_\max=4\times10^4$. For this given initial condition and 
$c=\sqrt{3}$, Eq.\ (\ref{degree}) becomes $D\le\sqrt{3}/\nu$. The viscosity 
$\nu=2.5/k_\max$ has been chosen in accord with this estimate 
to ensure that $k_\max$ lies well within the dissipation range. Our choice 
turns out to yield adequate dissipation, thus providing evidence for the 
sharpness of Eq.\ (\ref{degree}). We have used a standard 4th order 
Runge-Kutta method with the viscosity exactly incorporated through an 
integrating factor. The adapted time step $\delta t=-0.01/\sum_kku_k$ has 
been used to account for the highly sensitive nature of the problem when 
$t \approx t_*$.

\begin{figure} 
\begin{center}
\includegraphics[width=8.5cm]{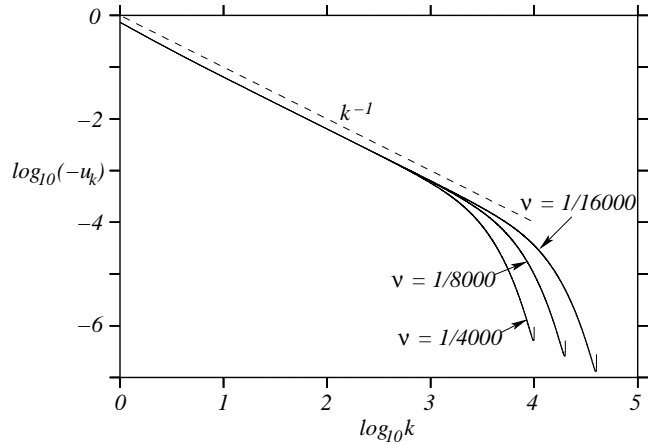}
\end{center}
\caption{Spectra ($-u_k$ vs $k$) at $t = \pi/2$ for the three smallest
values of viscosity considered, $\nu = 1/4000$, $1/8000$ and 
$1/16000$ (computed at resolutions $k_\max=10000$, $20000$ and $40000$
respectively).  Note, the spectra differ negligibly except in their
high wave number tails, and are well fit by a $k^{-1}$ slope in
the inertial range.}
\label{fig3}
\end{figure}

Figure \ref{fig3} shows the plots of $\log[-u_k(t_*)]$ versus 
$\log k$ for the three highest-resolution simulations. 
These exhibit a clear slope of $-1$ in 
the inertial range, thus implying the scaling $k^{-2}$ for the energy
spectrum. Evidently, the inertial range becomes wider for higher $\Re$, 
and a careful inspection of data also shows a clear trend that the 
inertial range becomes shallower, approaching the critical scaling 
$k^{-1}$ as expected.

\begin{figure} 
\begin{center}
\includegraphics[width=8.5cm]{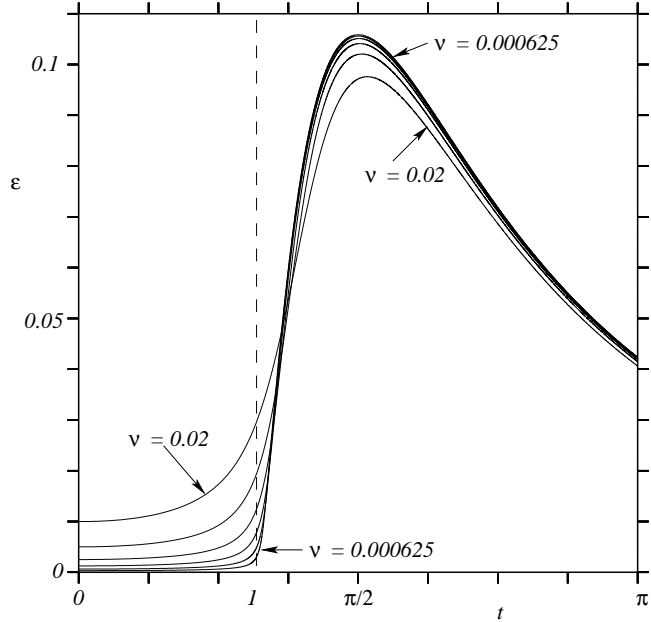}
\end{center}
\caption{Evolution of the energy dissipation rate $\epsilon(t)$ 
for a series of six simulations differing in $\nu$ by factors of 2 
(the extreme values of $\nu$ are indicated).  Also, the inviscid
singularity time ($t=1$) is indicated by the vertical dashed line.}
\label{fig4}
\end{figure}

\begin{figure} 
\begin{center}
\includegraphics[width=8.5cm]{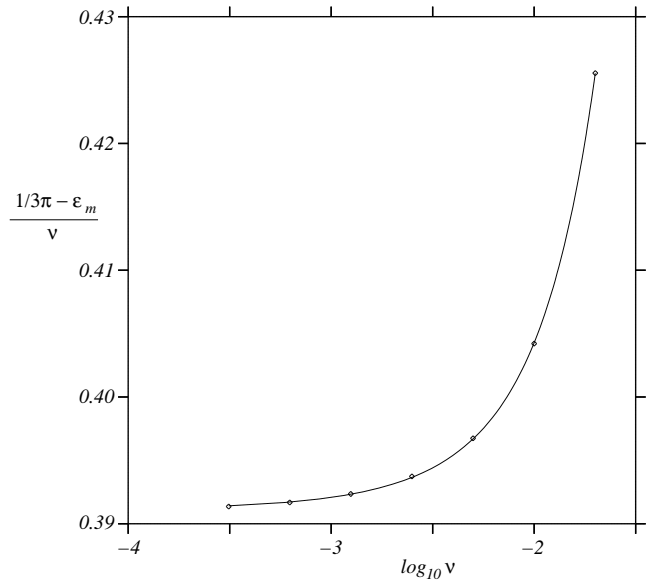}
\end{center}
\caption{least-squares quadratic fit 
$(1/(3\pi)-\epsilon_m)/\nu = 0.3911 + 0.9102\nu + 40.50\nu^2$ to
the numerical results indicated by the diamonds.}
\label{fig5}
\end{figure}

Figure \ref{fig4} shows the evolution of the energy dissipation 
rate $\epsilon(t)=\nu\norm{u_x}^2=\nu\sum_kk^2u_k^2/2$ 
from $t=0$ to $t=\pi$. The dissipation rate remains small for $t<1$
(evidently tending to zero in the inviscid limit), only to grow
considerably when $t=1$, consistent with the result (\ref{erate}) for
the limiting weak solution. This rate continues to increase for $t>1$
and achieves a maximum at $t=t_*=1.571$, which is very close to the
analytic value $\pi/2$. This value of $t_*$ has been observed to be
very robust with respect to independent variations of the Reynolds
number and the time step. The maximum dissipation rate is
$\epsilon_m=0.10605$ for the three highest Reynolds numbers.  This
suggests that the convergence of $\epsilon_m$ as $\nu\to0$ is rapid.
Indeed figure \ref{fig5} shows that $\epsilon_m$ differs only 
by approximately $0.39\nu$ from the theoretical limiting value
$1/(3\pi)$.  The curve in this figure shows the least-squares quadratic
fit $(1/(3\pi)-\epsilon_m)/\nu = 0.3911 + 0.9102\nu + 40.50\nu^2$ to
the numerical results indicated by the diamonds.

We now discuss the results from a second set of simulations, differing
from the first only in the initial condition: $u_2(0)=-1$ and $u_k(0)=0$ 
for $k\neq2$. In physical space this corresponds to $u(x,0)=-\sin2x$. 
For this case, only even wave numbers can be excited. Initially, the 
steepest slope is $-2$ occurring at $x=\pm\pi,0$, where the inviscid 
solution blows up simultaneously when $t=T=1/2$. One would expect 
$\epsilon_m$ to be twice as great as that in the previous case because 
the combined contribution to $\epsilon_m$ at both $x=-\pi$ and $x=\pi$ 
is equivalent to that at $x=0$. Furthermore, since the local extrema 
are $\pi/4$ away from the (stationary) locations of wave breaking, one 
would expect $t_*=\pi/4$. These are actually what we have observed. 
More precisely, the numerics have returned $\epsilon_m=0.2121$ and 
$t_*=0.7856$. The spectrum plot is the same as figure \ref{fig3} 
and is not shown.

In passing, it is worth mentioning that for the present example, 
$\epsilon_m$ can be made arbitrarily large by changing the initial 
condition. Given $u_\ell(0)=-1$ and $u_k(0)=0$ for $k\neq\ell$, which
corresponds to $u(x,0)=-\sin\ell x$ in physical space, only the wave 
numbers $\ell,~2\ell,~3\ell,\cdots$ can be excited. Initially, the 
steepest slope is $-\ell$ occurring at $x=2\pi n/\ell$ for 
$n=0,\pm 1,\pm2,\cdots$ and $|n|\le\ell/2$, where the inviscid solution 
blows up simultaneously when $t=T=1/\ell$. The local extrema are 
$\pi/(2\ell)$ away from the (stationary) locations of wave breaking.
One can expect $\epsilon_m=\ell/(3\pi)$ and $t_*=\pi/(2\ell)$, which 
we have actually observed (within small errors as the cases reported 
above) for several different values of $\ell$. Note that although 
$\epsilon_m$ can be made arbitrarily large by increasing $\ell$, Eq.\
(\ref{epsilonbound1}) does hold as both of its sides are proportional 
to $\ell$ ($k_0=\ell/L$). The scaling $E(k,t_*)=Ck^{-2}$, starting 
from $k=\ell$, has been observed to prevail for all cases, with 
$C\propto\ell$. 

\section{Conclusion}

In summary, we have studied both analytically and numerically 
one-dimensional viscous Burgers flows decaying from smooth initial 
conditions. The results obtained include upper bounds for the energy 
dissipation rate and number of degrees of freedom and constraints on 
the spectral distribution of energy. Given that the maximally achievable 
energy dissipation rate $\epsilon_m$ remains finite and positive in the
inviscid limit $\nu\to0$, it is found that energy spectra steeper than 
$k^{-2}$ are ruled out in that limit. For this critical scaling, 
$\epsilon_m$ satisfies
$\epsilon_m\le\sqrt{3}k_0\norm{u_0}_\infty\norm{u_0}^2$, where $k_0$ 
is the lower wave number end of the energy inertial range and $u_0$ 
is the initial velocity field. This further implies the upper bound 
$k_\nu\le\sqrt{3}\norm{u_0}_\infty/\nu$ for the energy dissipation 
wave number $k_\nu$. It follows that the number $N$ of Fourier modes 
within the energy inertial range satisfies 
$N\le\sqrt{3}L\norm{u_0}_\infty/\nu$, where $L$ is the domain size.
This result coincides with a rigorous estimate, using no assumption 
of power-law spectra, for the number of degrees of freedom $D$ defined 
in terms of local Lyapunov exponents.
 
As an illustrative example, we have considered both analytically and
numerically the Burgers equation in the periodic domain $[-\pi,\pi]$
with the initial condition $u_0(x)=-\sin x$. In the former approach,
we have tightened up the constraint on the spectral distribution of 
energy by pointing out that no power-law energy spectra other than
$k^{-2}$ are realizable. A detailed examination of the (unique) 
limiting weak solution has provided an explanation why the maximum 
velocity is better conserved than the energy. In the latter approach, 
we have demonstrated the exact $k^{-2}$ scaling and have numerically 
determined the viscosity-independent dissipation rate and time of 
maximum energy dissipation. These are consistent with analytic results 
derived from the limiting weak solution. 

\subsection*{References}

\noindent$^1$J. M. Burgers, ``A mathematical model illustrating the
theory of turbulence,'' Adv. Appl. Mech. {\bf 1}, 171 (1948).

\noindent$^2$E. Hopf, ``The partial differential equation 
$u_t+uu_x=\mu u_{xx},$'' Comm. Pure Appl. Math. {\bf 3}, 201 (1950).

\noindent$^3$J. D. Cole, ``On a quasi-linear parabolic equation occurring in
aerodynamics,'' Quart. Appl. Math. {\bf 9}, 225 (1951).

\noindent$^4$J. Bec and U. Frisch, ``Probability distribution functions of
derivatives and increments for decaying Burgers turbulence,'' Phys. Rev. E 
{\bf 61}, 1395 (2000).

\noindent$^5$J. Bec, U. Frisch, and K. Khanin, ``Kicked Burgers turbulence,'' 
J. Fluid Mech. {\bf 416}, 239 (2000).

\noindent$^6$A. Chekhlov and V. Yakhot, ``Kolmogorov turbulence in a 
random-force-driven Burgers equation,'' Phys. Rev. E {\bf 51}, R2739 
(1995).

\noindent$^7$W. E and E. Vanden Eijnden, ``On the statistical solution of
the Riemann equation and its implication on Burgers turbulence,'' Phys. 
Fluids {\bf 11}, 2149 (1999).

\noindent$^8$W. E and E. Vanden Eijnden, ``Another note on forced Burgers
turbulence,'' Phys. Fluids {\bf 12}, 149 (2000).

\noindent$^9$T. Gotoh and R. H. Kraichnan, ``Steady-state Burgers turbulence
with large-scale forcing,'' Phys. Fluids {\bf 10}, 2859 (1998).

\noindent$^{10}$V. H. Hoang and K. Khanin, ``Random Burgers equation and 
Lagrangian system in non-compact domains,'' Nonlinearity {\bf 16}, 819 (2003).

\noindent$^{11}$M. Vergassola, B. Dubrulle, U. Frisch, and A. Noullez, 
``Burgers-equation, devils staircases and the mass-distribution for
large-scale structures,'' Astron. Astrophys. {\bf 289}, 325 (1994).

\noindent$^{12}$V. Yakhot and A. Chekhlov, ``Algebraic tails of probability 
functions in the random-force-driven Burgers turbulence,'' 
Phys. Rev. Lett. {\bf 77}, 3118 (1996).

\noindent$^{13}$R. V. Y. Nguyen, M. Farge, D. Kolomensky, K. Schneider, and
N. Kingsbury, ``Wavelets meet Burgulence: CVS-filtered Burgers equation,'' 
Physica D {\bf 237}, 2151 (2008). 

\noindent$^{14}$C. V. Tran, ``Constraint on scalar diffusion anomaly in
three-dimensional flows having bounded velocity gradients,'' Phys. Fluids 
{\bf 20}, 077103 (2008).

\noindent$^{15}$J. Kaplan and J. Yorke, {\it Functional Differential 
Equations and Approximation of Fixed Points} 
(Springer, New York, p. 228, 1979). 

\noindent$^{16}$J. D. Farmer,  ``Chaotic attractors of an 
infinite-dimensional dynamical system,'' Physica D {\bf 4}, 366 (1982). 

\noindent$^{17}$C. V. Tran and L. Blackbourn, ``Number of degrees of
freedom of two-dimensional turbulence,'' Phys. Rev. E {\bf 79}, 056308 (2009).

\noindent$^{18}$C. V. Tran, ``The number of degrees of freedom of 
three-dimensional Navier--Stokes turbulence'' Phys. Fluids, in Press.

\noindent$^{19}$C. V. Tran, D. G. Dritschel, and R. K. Scott, ``Effective 
degrees of nonlinearity in a family of generalized models of 
two-dimensional turbulence,'' Phys. Rev. E, in Press. 

\noindent$^{20}$M. C. Cross and P. C. Hohenberg, ``Pattern formation
outside of equilibrium,'' Rev. Mod. Phys {\bf 65}, 851 (1993).

\noindent$^{21}$P. C. Hohenberg and B. I. Shraiman, ``Chaotic behaviour
of an extended system,'' Physica D {\bf 37}, 109 (1989).

\noindent$^{22}$C. V. Tran, T. G. Shepherd, H.-R. Cho, ``Extensivity of
two-dimensional turbulence,'' Physica D {\bf 192}, 187 (2004).

\noindent$^{23}$D. G. Dritschel, C. V. Tran, and R. K. Scott, ``Revisiting 
Batchelor theory of two-dimensional turbulence,'' J. Fluid Mech. {\bf 591}, 
379 (2007).

\end{document}